\documentclass[a4paper,11pt]{article}
\usepackage{boldline}
\usepackage{cancel}

\usepackage{cancel}
\usepackage{adjustbox}
\usepackage{amsmath}
\usepackage{amssymb}
\usepackage[T1]{fontenc}        
\usepackage{lmodern}    
\usepackage[utf8]{inputenc}     
\usepackage{ifthen}

\usepackage{graphicx}
\graphicspath{ {./images/}} 
\usepackage{adjustbox}

\usepackage{adjustbox}

\title{Fluctuating Latttice, Several Scales}
\author{H.B. Nielsen
  , Niels Bohr Institut\\
\vspace{-5mm}}

\date{Bled virtually, July  2024}

\begin{document}
\title{ Fluctuating Lattice, Several Energy Scales}
\author{
H.B.Nielsen, Niels Bohr Institute}
\maketitle
\begin{center}
  Part I: {\large Fluctuating Lattice, Relation between Scales}\\
  Part II: {\large Approximate Minimal $SU(5)$, Fine Structure Constants}
\end{center}

\begin{abstract}
  In part I: We find a series physical scales such as 1) Planck scale,
  2) Minimal
  approximate grand unification SU(5), 3) the mass scale of the see saw model
  right handed or Majorana neutrinoes, some invented scale with many scalar
  bosons, etc., and get the logarithms of these energy scales fitted by a
  quantity $q$ related to the dimensions of to the scales related
  dimensionalities of coefficients in Lagrangian densities, or some
  generalization of this $q$ to something similar in the various cases of
  the scales. The logarithm of the energies behave as a straight line
  versus the dimension related number $q$. This is being explained by
  an ontologically existing lattice, which fluctuates in lattice canstant
  $a$ fromplaceto place in space time or more precisely, it fluctuates
  quantum mechanically.
  
  In Part II: We find a fitting of the three fine structure constants in the
  Standard Model by means of a no-susy and SU(5)-like - but only accurately 
  SU(5) symmetric in the classical approximation - by means of three other
  parameters for each of which, however, we have speculative predictions:
  Quantum corrections due to the lattice which are three times as large as
  naive quantum corrections, because the lattice is supposed to lie in
  layers, one layer for each fermion-family; criticallity of the unified
  coupling, using the satnadrd model group $S(U(2)\times U(3))$; the
  unification scale is fitted inot the scale-system of part I. 
  \end{abstract}

   %
%

\section{Overall Introduction}

This article is composed of two parts, the first(review of \cite{RSR}),
in section \ref{sI}, of
which fits a series
of energy scales, or we could say physical phenomena leading to a parameter
(with dimension energy after we put $c=\hbar=1$) and thus potentially a/the
``fundamental'' energy scale, while the other part(review of \cite{AppSU5})
section \ref{sII} could
be considered
an attempt to ``rescue'' grand unification SU(5) by interpreting it by being
only an (accidental) classical approximation.

The overall plan will be so that we first discuss the many energy
scales, in section \ref{sI},
which we seek to unite by means of the model,  which is a really existing
lattice fluctuating in size.

Next comes the rescue of the Grand Unification\cite{GG, WGG3,GG2} in
section  \ref{sII} by
allowing it to be
only a classical approximation\cite{AppSU5,Senjanovic}; but then after that
we return in section
\ref{sp} to the
attempt to unify the energy scales in the light, that it is really very
much needed, because giving up the usually used susy\cite{LD,LRSU5,Enkhbat} to
make SU(5) GUT
work points to lower unification energy scale than even the susy SU(5)
unification, so that now the distance even in logarithm between the
Planck scale and the unificationscale has become so large that it is hard to
see how to make them compatible. Thus the call for our attempt to unify the
scales by our fluctuating lattice\cite{RSR} has got even more strong, than with
usual susy unification.

The spirit of the present work is the attempt to get information about the
underlying theory to be found by numerological scuccess from the information
in the Standard Model parameters as counted in \cite{Rugh} or in other
observations the right handed neutrino masses from neutrino-oscillations
\cite{seesaw} or inflation \cite{anintr, Enquist}.

\section{Introduction for first part: Energy Scales or Fluctuating lattice}
\label{sI}
\begin{frame}

  Several seemingly ``fundamental'' scales like the Planck
  scale, the see-saw scale\cite{seesaw}, and the unification scale, are not so
  equal to each other as we would have expected in a philosophy of their
  being only {\bf one} ``fundamental energy scale''.

  To cure this fact we bring forward the idea of a truly existing
  lattice, which has {\bf wildly different sizes}, of say the link
  length,  in different places and/or in different components of a
  superposition. Really we think of the lattice as being superpositon
  of all possible deformations (which could be made by coordinate
  transformations/reparametrizations), so we can say we have in mind
  quantum fluctuations in reparametrizations, the gauge group of gravity.

\end{frame}

In the tables in which we list the various energy scales  
we have attached to each scale a number $q$ or $n=4-q$, and
the meaning is that $n$ should mean the power to which the inverse
link size $1/a$ be raised in order to give weighting comming in in the
calculation of the energy scale in question. In fact this means that for all
the energy scales we have
\begin{eqnarray}
  \hbox{``energy scale''}&\approx& \sqrt[n]{<(1/a)^n>}
\end{eqnarray}
If the distribution in the fluctuating lattice had been very norrow this
$n$ would make no difference, but assume and fit a very broad distrbution
being a Gaussian distribution in $\ln a$,
\begin{eqnarray}
  P(\ln a)d\ln a&=& \frac{1}{2\pi \sigma}
  \exp(-\frac{1}{2\sigma}(\ln(a)-\ln(a_{0\; counting})^2)d\ln(a)\label{distrib}
  \end{eqnarray}
where the spread $\sigma=5.5$ turns out\cite{RSR}, a rather large spread, when
we have in mind that it is a spread in the logarithm. In first
approximation we might think of a completely flat distribution in the logarithm.
If there was scalings symmetry and we got the Haar measure distribution
for the group of scalings with different factors, then we would get such a
flat distribtuion in the logarithm $\ln(a)$. Actually you might think of
(\ref{distrib}) as a scalings symmetric distribution only weakly broken.

For the purpose of comparing with the $P_{vloume}(\ln(a))$ below we might write
uding that density of hypercubes per unit  four volume compared the one per
density per four volume of a hypercube in the lattice is
$1/a_{0\; counting}^4$. Using that we can write for the number of hypercubes
in an infinitesimal region
\begin{eqnarray}
  P(\ln a)d\ln a&=& \frac{\frac{1}{2\pi \sigma}
  \exp(-\frac{1}{2\sigma}(\ln(a)-\ln(a_{0\; counting})^2)}{a_{0\; counting}^4}d^4xd
  \ln(a)\label{distrib2}.
  \end{eqnarray}

\subsection{Density definitions etc.}

First let us be a bit more  specific about how we think of this
fluctuating lattice:
\begin{itemize}
\item{\bf Lattice in layers} We imagine, that the lattice can
  lie in several layers compared to a simple Wilson lattice.
  An  idea about having layers is best gotten by imagining, that
  we take a number , e.g. 3, Wilson lattices and have in nature all
  of them. Then for each fourvolume of the size of a hypercube
  in the lattice in space time we shall not have as in a simple
  Wilson lattice just one lattice site, but rather 3. We call this, that there
  are 3 layers. If the lattice fluctuate in size of the links and thus
  this size also varies of course from place to place in the
  Minkowski space time, we can in principle ask for such a number of
  layers in average for each size of link $a$. That is to say
  we can define a ``numbers of layers for a small $a$ region '' = `` numbers of
  sites(or hypercubes) in the four volume of one single hypercube
  provided we count only hypercubes in a range of sizes $a$ given by
  say the infinitesimal $d\ln(a)$''
  \begin{eqnarray}
    P_{layer}(\ln(a)) d\ln(a)&=& \hbox{`` numbers of hyper cubes with
      link-size}\nonumber\\
    && a'\in\{a'|a\le a' \le a*\exp(d\ln(a))\} \hbox{ per four}\nonumber\\
    &&  \hbox{ volume $a^4$ of the hypercubes for $a$''}\\
    &=& \hbox{``Layer density''($a$)}d\ln(a) 
    \end{eqnarray}
\item{\bf Density in Space Time} Usually we consider of course densities
  per 4-volume of Minkowski (or the curved space time) space and then there is
  place in every layer for having $1/a^4$ four-cubes. So if the density of
  say sites per place for a four cube in one of layers say is per interval
  in the logarithm $d\ln(a)$ is $\hbox{``Layer density''}(a) d\ln(a)$ (and there are $1/a^4$ per $(unit\; length)^4$),
  then the density of sites per four volume in Minkowski space time is \begin{equation}
  \def\arraystretch{1.1}
  \begin{array}{r@{\;}l} 
    P_{volume}(\ln(a))d^4xd\ln(a)&= \hbox{``Layer density''}(a)/a^4
    d^4x d\ln(a)\\
      &= P_{layer}(\ln(a))/a^4d^4xd\ln(a)\\
      \hbox{with ansatz (\ref{distrib})}: \;\;&= \frac{1}{2\pi \sigma}
      \exp(-\frac{1}{2\sigma}(\ln(a)-\ln(a_{0\; counting})^2)/a^4d^4xd\ln(a)\\
      &=  \frac{1}{2\pi \sigma}
      \exp(
      -\frac{1}{2\sigma}(\ln(a)-\ln(a_{0\; counting})+4\sigma)^2+\nonumber\\
      &+4^2/2*\sigma
      -4\ln(a_{0\; counting}) )d^4xd\ln(a)\\        
\text{Calling} 
\ln(a_0) &=\ln(a_{0 \; counting})-4\sigma\\
      \hbox{or } a_0&= a_{0\; counting}\exp(-4\sigma) \\
      \hbox{then }\\ P_{volume}(\ln(a))d^4xd\ln(a)&=  \frac{1}{2\pi \sigma}
      \exp(
      -\frac{1}{2\sigma}(\ln(a)-\ln(a_0))^2- \nonumber\\
      &-4^2\sigma/2
       -4\ln(a_{0}) )d^4xd\ln(a)\\
       &= \frac{1}{2\pi \sigma}\exp(-\frac{1}{2\sigma}(\ln(a)-\ln(a_0))^2-
       \nonumber\\
       & -8\sigma-4\ln(a_0))d^4xd\ln(a)\\
    &=  \frac{1}{2\pi \sigma}\exp(-\frac{1}{2\sigma}(\ln(a)-\ln(a_0))^2)
       /a_0^4*\exp(-8 \sigma)d^4xd\ln(a)\nonumber\\
     &=  \frac{1}{2\pi \sigma}\exp(-\frac{1}{2\sigma}(\ln(a)-\ln(a_0))^2)
       /a_{0\; counting}^2/a_0^2 d^4xd\ln(a)
  \end{array}
\end{equation}

  But if we should normalize properly this $P_{volume}(\ln(a))$ properly we
  should have had like in (\ref{distrib2}) that the division with
  $a_{0\;counting}^2a_0^2$ should be replaced by $a_{0\; counting}^4$. Thus the
  normalized $P_{volume}(\ln(a))$ looks rather
  \begin{eqnarray}
    P_{volume}^{(N)}(\ln(a))d\ln(a)d^4x/a_{0\; counting}^4&=&
    \frac{a_{0\; counting}^2}{a_0^2}P_{volume}(\ln(a))\nonumber\\
    && d\ln(a)d^4x/a_{0\; counting}^4\\
    &=& \exp(8\sigma)P_{volume}(\ln(a))\nonumber\\
     && d\ln(a)d^4x/a_{0\; counting}^4
    \end{eqnarray}
  \end{itemize}
  This means that when we average over the distribution of the link length
  $a$ or this link length $a$ to some power $a^p$ say the answer is not the
  same as if we do it just averaging over links or hypercubes by their number.
  In fact denoting the two different averages $<..>_{counting} $ and
  $<...>_{volume}$ we find
  \begin{eqnarray}
    <a^p>_{counting} &=&\int \frac{1}{2\pi\sigma}\exp(-\frac{1}{2\sigma}
    (\ln(a) - \ln(a_{0\; counting}))^2) * a^p d\ln(a)\\
    &=& \int \frac{1}{2\pi\sigma}\exp(-\frac{1}{2\sigma}
     ((\ln(a) - \ln(a_{0\; counting}))^2-2p\sigma*\ln(a) )) d\ln(a)\nonumber\\
    &=&\int \frac{1}{2\pi\sigma}\exp(-\frac{1}{2\sigma}
     (\ln(a) - \ln(a_{0\; counting}-\sigma*p)^2 +\nonumber\\
    &&+p^2\sigma/2 +
    p * \ln(a_{0\; counting}) ) d\ln(a)\\
    &=& a_{0\; couning}^p\exp(p^2\sigma/2)\\
    \hbox{so that } \sqrt[p]{<a^p>_{counting}} &=& a_{0\; counting}\exp(p\sigma/2)\\
    \hbox{while } <a^p>_{volume} &=&
    a_0^p*\exp(p^2\sigma/2)\\
    \hbox{so that } \sqrt[p]{<a^p>_{volume}}&=& a_0\exp(p\sigma/2).\\
    \hbox{Of course } <a^p>_{volume}& =& \frac{<a^p/a^4>_{counting}}
         {<a^{-4}>_{counting}}\\
    \hbox{as checked:} &=& \frac{a_{0\; counting}^{p-4}\exp((p-4)^2\sigma/2)}
         {a_{0\; counting}^{-4}\exp((-4)^2\sigma/2)}\\
         &=& a_{0\; counting}^p\exp(p(p-8)\sigma/2)\\
         &=& a_0^p\exp(p^2\sigma/2).\\
           \hbox{Also:}\frac{\langle a^{p+b}\rangle_{counting}}
                {\langle a^p\rangle_{counting}}&=&
                a_{0\; counting}^b\exp((p+b)^2-p^2)\sigma/2)\\
           &=& a_{0\; counting}^b\exp(b(2p+b)\sigma/2)\\
           \hbox{so that } \sqrt[b]{\frac{<a^{p+b}>_{counting}}{<a^p>_{counting}}}
           &=& a_{0\, counting}\exp(2p\sigma/2)* \exp(b\sigma/2)\\
           &=& a_{0\; p}\exp(b\sigma/2)\\
           &=&\sqrt{a_{0\; p}a_{0\;p+b}}\\
           \hbox{where } a_{0 \; p}&=& \hbox{Peak of } P_{layer}(\ln(a))*a^p\\
           &=& \hbox{Peak of }\exp(-\frac{1}{2\sigma}(\ln(a) -
           \ln(a_{0\;counting})^2+p\ln(a))\\
           &=& a_{0 \; counting}\exp(p\sigma)\\
           \hbox{so especially }a_0 &=& a_{0\; p=-4}=a_{0\; counting}\exp(-4\sigma)\\
    \end{eqnarray}
  

  Let us especially learn that considering two averages of powers
  of the link variable in succession you get an increase by a factor
  \begin{eqnarray}
    \frac{<a^{p+1}>_{counting}}{<a^p>_{counting}}&=& a_{0\; counting}
    \exp((p+1/2)\sigma).
  \end{eqnarray}
  So the effective energy scale when you work with powers of the
  link variable of the order of $p$ is about $a_{0\; counting}*\exp(p\sigma)$.
  So what we have to do to evaluate what the typicalpower is for the
  type of physics connected to the energy scale we want.

  Basically our procedure is to represent the quantity, which we call the
  energy scale, as a root of or just the coefficient in the action or a
  ratio of such action related quantities, and then argue that this combination
  must - assuming no big (or small) numbers in other coupling or parmeters -
  that it should behave as the average of some power of the link length $a$,
  i.e. as say $<(1/a)^n>_{counting}= <a^{-n}>_{counting}$.

  As just a repetition we used in the tables below also a to the power
  $n$ of $1/a$ equivalent number $q=4-n$,a notation inspired by
  considering the scales ``see-saw'' and ``scalars'' which are scales at
  which we postulate/speculate, that there are ``a lot of'' respectively
  fermion and boson masses. In fact we know that ignoring the interactions for
  simplicity the fermion and boson actions in field theory are
  \begin{eqnarray}
    S_{fermion}=\int {\cal L}_D(x)d^4x&=&\int \bar{\psi}(x)
    (i\gamma^{\mu}\partial_{\mu}
    -m)\psi(x) d^4x+...\\
    S_{scalar \; boson} = \int {\cal L}(x)d^4x &=& \int  ( \frac{1}{2}
    \eta^{\mu\nu}\partial_{\mu}\phi\partial_{\nu}\phi- \frac{1}{2}m^2\phi^2
     )d^4x+...
    \end{eqnarray}
  and that the mass $m$ of the particle occurs in different
  powers $m^q$ for the two, namely $q=1$ forfermions and
  $q=2$ for bosons. For dimensional
  reasons these mass terms then - including the extra $1/a^4$ factor
  from number of hyper cubes going as $1/a^4$ in a unit space time -
  have link $a$ dependensies
  \begin{eqnarray}
    \hbox{mass term $a$ dependence } &\propto& (1/a)^{4-q}.
    \end{eqnarray}
  and so we have here $n=4-q$.

  (Unfortunately we have in Part II used the letter $q$ for a non-integer
  quantity being the parameter for how much the finestructure constant
  deviate from the SU(5) prediction, which we predict to $q=3\pi/2$)
  
\vspace{1cm}
  
  \subsection{Our tables}

  Let us first deliver the table of the energy scales I included in
  the very workshop talk:

  When we have to do with the quantities related to the scale being
  terms in lagrangians in a field theory, we have immediately a factor
  $1/a^4$ which means a $-4$ in the power.

  It shall turn out from our fitting that the step in the energy scale
  per unit step in the power is a factor 251, which must then be identified
  with our step factor $\exp(\sigma)$.

  \vspace{3cm}
  
  {\bf Table of ``Fundamental ?'' Energy Scales}
  
\begin{adjustbox}{width=\textwidth, center}
  \begin{tabular}{|c|c|c|c|c|c|c|}
    \hline
    Name&Energy value&$n$ of $(1/a)^n$&q&Coef. dim.& Fit& Lagrangian d.\\
    \hline
  Planck scale& $1.22*10^{19}$&6&-2&-2&$2.44*10^{18}$GeV&
   $\frac{1}{2\kappa} R$\\
  reduced Planck& $2.43*10^{18}$GeV&6&-2&-2&$2.44*10^{18}$GeV&
  $\frac{1}{2\kappa} R$\\
  \hline
  Min. $SU(5)$ app.& $5.3*10^{13}$GeV&4&0&0&$3.91*10^{13}$ GeV&
  $-\frac{1}{16\pi\alpha}F_{\mu\nu}^2$\\
  Susy SU(5)& $10^{16}$ GeV&4&0&0&$3.91*10^{13}$ GeV
  &-$\frac{1}{16\alpha}*F_{\mu\nu}^2$ \\
  \hline
  See-saw& $10^{11}$ GeV& 3&1&1&$1.56*10^{11}$GeV&$m_R \overline{\psi}
  \psi $\\
  \hline
  Fermion extrapolate& $10^4$GeV &0&4&4&$10^4$ GeV&``1''\\
  \hline
\end{tabular}
\end{adjustbox}

\vspace{1cm}

Using $c=\hbar =1$,
\begin{eqnarray}
  \hbox{``Reduced Planck''} &=& \frac{1}{\sqrt{G_{Newton}*8\pi}}
  =2.43*10^{18}GeV\label{ReducedPlanck}
  \\
    \hbox{``unified (approximate) SU(5)''}&=&
    \hbox{``where lines closest''}
    \nonumber
    \\
    & =&(say) 5.3*10^{13}GeV\label{unification}
    \\
    \hbox{``see-saw''} &\sim& \hbox{``typical right handed neutrino mass''}
    \nonumber\\
    &\sim& 10^{11}GeV.\label{seesaw}
      \end{eqnarray}

\vspace{1cm}

The `scales'' {\bf scalars} and {\bf Fermion extrapolate}
also called ``fermion tip'' scale) are my inventions
and need explanation later. ($\kappa=8\pi G= 8\pi G_{Newton}$).

Next let me deliver the full table (from \cite{RSR}) with scales which
we now have found,
including cases of scales judged by very good will:

\vspace{8mm}

{\bf Important to notice in this table is the very good fitting of
  our formula}
\begin{eqnarray}
  \hbox{``energy scale''}&=& 10^4 GeV * 250^n
\end{eqnarray}
{\bf which is compared with the in the third column of the table.
The fitted value is the lower one inside each block.}

\begin{adjustbox}{width =\textwidth}
\begin{tabular}{|c|c|c|c|c|}
  \hline
  Name&[Coefficient]&``Measured'' value&Text ref.&$n$ \\
  comming from&Eff. $q$ in $m^q$ term& Our Fitted value&Lagangian dens.&by\\
  status&&&&$1/a^n$\\
  \hline
  \hline
  Planck scale&$[mass^{2}]$ in kin.t.&$1.22*10^{19}GeV$&
  &6\\
  Gavitational $G$&q=-2&$2.44*10^{18}GeV$&$\frac{R}{2\kappa} $&\\
  wellknown&&&$\kappa= 8\pi G $&\\
  \hline
  Redused Planck&$[mass^{2}]$ in kin.t.&$2.43*10^{18}GeV$&(\ref{ReducedPlanck})&6\\
  Gravitational $8\pi G$&q=-2&$2.44*10^{18}GeV$&$\frac{R}{2\kappa}$&\\
  wellknown&&&$\kappa=8\pi G$&\\
  \hline\hline
  Minimal $SU(5)$&$[1]$&$5.3*10^{13}GeV$&(\ref{unification})&4\\
  fine structure const.s $\alpha_i$&q=0&$3.91*10^{13}GeV$&
  $\frac{F^2}{16\pi \alpha}$&\\
  only approximate&&&$F_{\mu\nu}=\partial_{\mu}A_{\nu}-\partial_{\nu}A_{d\mu}$&\\
  \hline
  Susy $SU(5)$&$[1]$&$10^{16}GeV$&
  &4\\
  fine structure const.s&q=0&$3.91*10^{13}GeV$&$\frac{F^2}{16\pi \alpha}$&\\
  works&&&$F_{\mu\nu}=\partial_{\mu}A_{\nu}-\partial_{\nu}A_{\mu}$&\\
  \hline
  \hline
  Inflation $H$&$[1] ?$&$10^{14}GeV$&
  &4\\
  CMB, cosmology&q=0?&$3.91*10^{13}GeV$&$\lambda \phi^4$&\\
  ``typical'' number&&&$V=\lambda \phi^4$&\\
  \hline
  Inflation $V^{1/4}$ &concistence ?&$10^{16}GeV$  &
  &5\\
  CMB, cosmology&q=-1?&
  $9.96*10^{15}GeV$&consistency&\\
  ``typical''&&&$V=\lambda \phi^4$?&\\
  \hline
  \hline
  See-saw& $[mass] in \; non-kin.$ &$10^{11} GeV$& (\ref{seesaw})&3\\
  Neutrino oscillations& q=1& $1.56*10^{11}GeV$&$m_R\bar{\psi}\psi$&\\
  modeldependent& &&$m_R$ right hand mass&\\
  \hline\hline
  Scalars&$[mass^2] in\; non-kin.$&$\frac{seesaw}{44\; to\; 560}$&
  &2\\
&&&&\\
  small hierarchy& q=2&$\frac{1.56GeV}{250}$& $m_{sc}^2|\phi|^2$&\\
  invented by me&&&breaking $\frac{seesaw}{scalars}$&\\
  \hline\hline
  Fermion tip& $``[mass^4] in\; non-kin.''$& $10^4 GeV$&
  &0\\
  fermion masses& q=4& $10^4GeV$& ``1''&\\
  extrapolation &&&quadrat fit&\\
  \hline\hline
  Monopole& $``[mass^5] in\; non-kin.''$& $28 GeV$&
  &-1\\
  dimuon 28 GeV& q=5 & $40 GeV$&$m_{monopol}\int ds$&\\
  invented&&&$S \propto a$&\\
  \hline\hline
  String $1/\alpha'$ & $``[mass^6] in\; non-kin.''$& $1 Gev$&
  &-2 \\
  hadrons& q=6& $0.16 GeV$&Nambu Goto&\\
  intriguing &&&$S\propto a^2$&\\
  \hline
  String $T_{hagedorn}$& $`[mass^6]$ in\; non-kin.''& $0.170GeV$&
  &-2\\
  hadrons&q=6& $0.16 GeV$ & Nambu Goto&\\
  intriguing&&&&\\
  \hline\hline
  Domain wall&``$[mass^7]$  in \; non-kin.''&$8MeV$&
  &-3\\
  dark matter&q=7&$0.64 MeV$&twobrane vol.&\\
  far out&&&&\\
  \hline \hline
\end{tabular}
\end{adjustbox}

\

After the Bled I found several new scales that in fact quite remarkably
fitted in rather well. In the recent articel  	arXiv:2411.03552 [hep-ph]
\cite{RSR}
I extended the table of scales fitting to by the scales ``monopoles'' and
``strings'' for which we what would be with the fluctuating lattice
the scales for monopole masses and for the string tension or say
the Regge slope in the Veneziano model, when one estimate the actions for
a single monopole or for a single string in our scheme. Quite surprisingly
the string scale turning out is the one for hadronic strings with which
historically string theory were first proposed. The monopole mass scale
turns out not so far from the mass of two-muon resonance\cite{dimuon} with
28 GeV mass, which is one of the very few peaks found not belonging
to the Standard Model in LHC.

Our fitting of the curve of scales by a linear function as function
of the power $n$ may be presented as
\begin{eqnarray}
  \hbox{``energy scale''} &=& 10^4 GeV * 250^n\\
  \hbox{or }  log_{GeV} (\hbox{``energy scale''}) &=& 4+2.40*n (+log GeV)\\
  \hbox{or } \ln_{GeV}(\hbox{``energy scale''}) &=& 9.21+ 5.53*n (+\ln GeV)
\end{eqnarray}
%
  {\bf Different  ``Energy-scales'' versus n = 4 -$(\pm)$``Coupling dimension''}
  
  \includegraphics[scale=0.5]{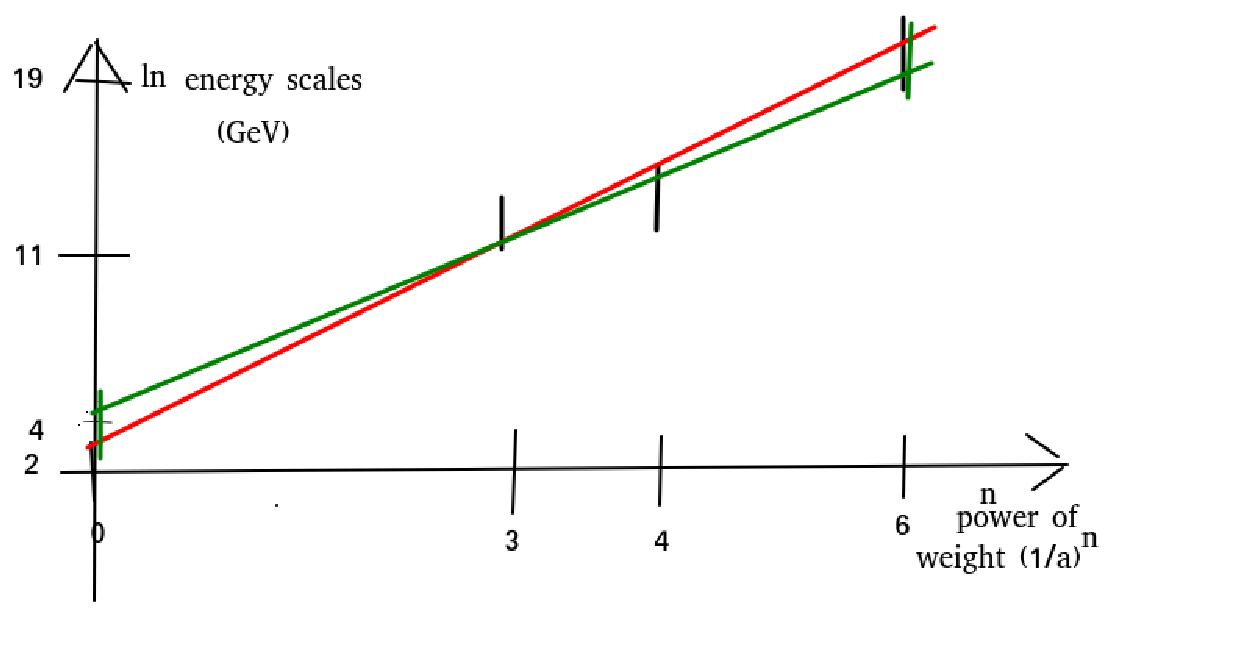}
  
\begin{frame}

  {\bf If we include names for the scales on the figure, it looks:}
  
  \includegraphics[scale=0.5]{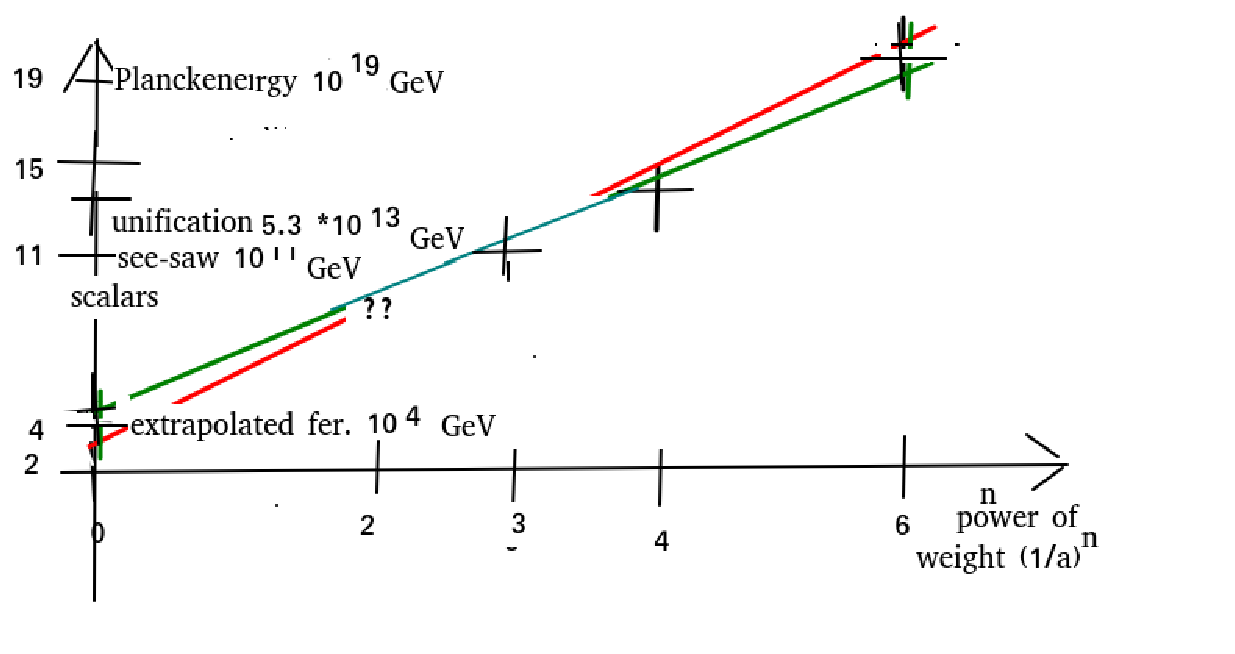}
  \end{frame}
\begin{frame}

  \begin{figure}
    \includegraphics[scale=0.87]{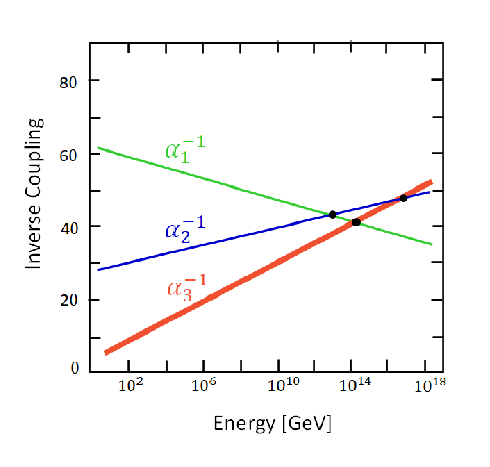}
    \caption{\label{Running}  {\bf If I do not believe Susy, just look
        where lines closest}. This is the usual plot of the running inverse
      fine structure constants normalized with the usual 3/5 on the
      U(1) coupling, so that they should have met in one point all three
      if exact SU(5) GUT had worked. But now we here refuse to believe
      in susy and instead assume the SU(5) symmetry only to be an approximate
      symmetry (see more in Part II) we take the energy (which is
      on the abscissa) at which the the three inverse fine structure constant
      curves are the closest to be the approximate unification scale.
      In our work \cite{AppSU5} we found by more accurate lattice model
      assumptions $5.3*10^{13}GeV$ for the approximate unification scale.
      This approximate unification scale is indicated on the figure by
      a vertical line.}
  \end{figure}
\end{frame}

\begin{frame}
  
  {\bf My Speculations on GUT SU(5): Approximate}
  \begin{itemize}
  \item There is a physically existing lattice, and the Plaquette
    action happens {\bf classically} to be SU(5) symmetric.
  \item Quantum corrections break the classical SU(5) symmetry of the
    lattice action.
  \item Because the lattice lies in three layers, the quantum correction SU(5)
    breaking is just a factor 3 bigger than true quantum correction.
    The factor 3 is the number of families.
    \end{itemize}
\end{frame}

We shall return to this model of an only approximate grand unification
SU(5) in the second part \ref{sII}.

\begin{frame}
  
  {\bf Take at First that our Energy Scales (4 of them, when I gave the talk;
    now $\approx$ 9) are Observed
    Phenomenologically on Logarithmic Plot as function of the Coupling Constant
    Dimension $[GeV^{dim}]$ a bit modified to be a power of $1/a$
    expected relevant lie on Straight line}

  Since the power $n$ to which the inverse link length $a$ comes into the
  action $S$ for the Lagrangian densities for the different sort of physics
  related to the different (fundamental?) energy scales, $(1/a)^n$, is
  linearly related to the dimension of the coefficient [``coefficient'']
  \begin{eqnarray}
    n&=& 4 - (\pm)Dim(``Coefficient''),
  \end{eqnarray}
  linearity - i.e. straight line - of the logarithm of the energy scales
  as function of $n$ means also, that we {\bf observed straihgt line
    for the relation of Dim(``coefficient'') to logarithm(``energy scale'')}
\end{frame}

\subsection{Fluctuating Lattice}

\begin{frame}
  
  {\bf ``Fluctuating lattice'' (in superposition of) being dense somewhere
    and rough somewhere and often deformed}
  \begin{figure}
    \includegraphics[scale=0.6]{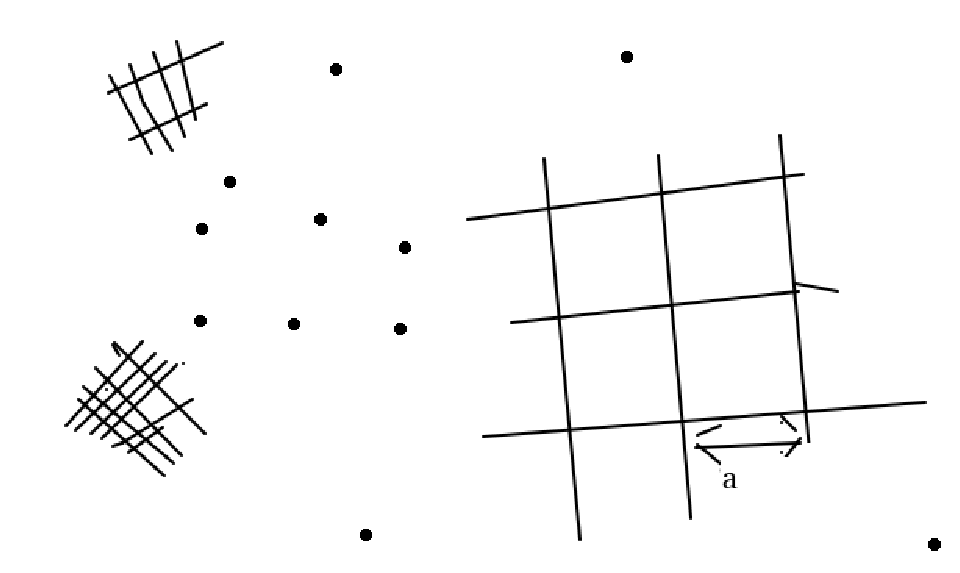}
    \caption{\label{fl} Here we drew for simplicity and easiness
      a couple of seperate pieces of lattices (two dimensional but
      of course they should really have been 4 dimensional to be in Minkowski
      space) of different lattice constant $a$ length and thus density.
      Meant is, however, that the lattice hangs together and that the
      link length varies along as one goes fromplace to place, presumably
      in a continuous way. Raelly it quantum fluctuates.}
  \end{figure}

\end{frame}
\begin{frame}
  
  {\bf Densities of, say, Sites in Fluctuating Lattice with several Layers
    can be defined relative to the hypercube four-volume $a^4$ or with
  respect to a fourvolume unit $m^4$ say:}

  For the distribution of different densities of links or of sites in a
  fluctuating lattice with shall distinguish:
  \begin{eqnarray}
    \# layers &=& (say) \; \# sites\; \hbox{per } a^4\\
    density_{/m^4} &=& \# sites\;  \hbox{per }m^4\\
    density_{/m^4}&=& \# layers *a^{-4}\\
    \hbox{Ansatz:}&&\nonumber\\
    \hbox{Probability: }&&\\
    P(\ln (1/a)) d\ln(1/a) d^4x &\propto&\nonumber\\
    \propto 
      \exp(-\frac{(\ln(1/a) - \ln(10^4 GeV))^2}{2\sigma})&&d\ln(1/a) d^4x
  \end{eqnarray}
  
  where $10^4GeV$ is our ``fitted'' value; $\sigma$ is a spreading to be
  fitted.
  \end{frame}
\begin{frame}
  
  \begin{center}
  {\bf Distribution of Contribution for one of the to Scales associated
    actions versu $\ln(a)$}
  \end{center}
  
  One of the actions associated with the candidates for fundamental
  scales as e.g. the Einstein-Hilbert-action $\frac{1}{2\kappa}R\sqrt{-g}d^4x$
  with $(1/a)^n$ proportional contribtuion get of the form:

  \begin{eqnarray}
    S =&&
    \int {\cal L}(x) d^4x\propto \nonumber\\
    &\propto& \int
    \exp(\frac{-(\ln(1/a)) - \ln(10^4GeV))^2}{2\sigma}) (1/a)^nd\ln(1/a)
    \nonumber\\
    &=& \exp(\frac{-(\ln(1/a) - \ln(10^4GeV))^2+n*2\sigma*\ln(1/a)}{2\sigma})
    d\ln(1/a)
    \nonumber
    \end{eqnarray}
  \end{frame}

\begin{frame}

  \begin{center}
  {\bf An Action depends on the spread $\sigma$ like:}
  \end{center}
  
  \begin{eqnarray}
  S&=& \int {\cal L}(x) d^4x\propto\nonumber \\
  &\propto&\int
  \exp(-\frac{(\ln(1/a))-\ln(10^4 GeV) -n*\sigma)^2+(n*\sigma)^2}
      {2\sigma})d\ln(1/a)
    \nonumber\\
    &=& \int
  \exp(-\frac{(\ln(1/a))-\ln(10^4 GeV))^2+(n*\sigma)^2}{2\sigma})d\ln(1/a)
    \nonumber\\
    &=& \int
  \exp(-\frac{(\ln(1/a))-\ln(10^4 GeV))^2}{2\sigma})d\ln(1/a)*\exp(n^2\sigma/2)
    \nonumber
  \end{eqnarray}
  where only the last factor $\exp(n^2*\sigma/2)$ depends on $n$.
  This was for an action $S  \propto (1/a)^n$.

  Interpreting the factor $\exp(n^2\sigma/2)$ as correcting the $n$ factors
  $(1/a)$ we get $(1/a)_{eff} =
  (1/a) *\exp(n\sigma
  )$.(because a step in $n^2\sigma/2$ is say
  $(n+1)^2\sigma/2 - n^2\sigma/2 = (2n+1)\sigma/2 \approx n\sigma$).
  \end{frame}
\begin{frame}
  
  {\bf The Effect of the in $\ln(1/a)$ Broadened Distribution is $1/a
    \rightarrow (1/a)_{eff}=\exp(n\sigma
    )*(1/a)$}

  We shall interprete correction to the effective $1/a$ (= the inverse of the
  link size) as a correction of the ``energy scale''. So the effect of the
  spreading with Gauss distribution in the logarithm $\ln(1/a)$ with a width
  given by $\sigma$ as
  \begin{eqnarray}
    \hbox{Replace }\hbox{``energy scale''}&\rightarrow& \hbox{``energy scale''}
    *\exp(n*\sigma
    )\\
    \hbox{So } 250 &=& \exp(\sigma/2)
    \hbox{(where $250$ from our empirical fit)}\nonumber\\
    \hbox{and thus } \sigma &=&
    5.5. 
    \end{eqnarray}
  \end{frame}

\subsection{Conclusion of Energy Scales and Fluctuating Lattice
  Part/First Part }
\begin{frame}
  
  \begin{itemize}
  \item We presented an emprical straight line fit to three wellknown
    energy scales, valid to crude order of magnitude accuracy,
    \begin{eqnarray}
      \hbox{``energy scale''}&=& 10^4GeV* 250^n\\
      \hbox{or }\hbox{``energy scale''}&=& 10^4GeV*250^{4-(\pm)dim(coefficient)}
    \end{eqnarray}
    (where $dim(coefficient)$ is the dimension in energy units of the
    coefficient multiplying in the Lagrangian density the field (product),
    and $(\pm) =+1$ for the term with the coefficient being a mass term
    like in the case of the ``see-saw''and the ``scalars scale'', while
    in the case of ``Plack scale'' where it iis the Einstein Hilbert action,
    which is a kinetic term carrying a dimension 2 coefficient $(\pm) =-1$
    )
  \item We explain this empirical fit with a speculated ``fluctuating
    lattice'' with a fluctuation distribution being a Gauss distribultion
    in the logarithm of the statistically fluctuating link length $a$, i.e.
    a Gauss distribution in $\ln(1/a)$:
    \begin{eqnarray}
      P&\propto& \exp(-\frac{(\ln(1/a)-\ln(10^4GeV)^2}{2*
      5.5}).
      \end{eqnarray}
    \end{itemize}
\end{frame}

\subsubsection{Last moment development}
After the talk I found several new ideas for new scales (mainly) of the type
that one considers a brane of some dimension $D$, meaning a space time
extension of dimension $D+1$, and thus having an action, which using
dimensionless parameters to makecoordinates on the brane-time-track,
acoefficient of dimension like $a^{D+1}$ being $[energy^{-D-1}]$, so that
$d=-D-1$. Very interesting it seems that for a string = (D=1)brane the
tension of the strings pointed to by our extrapolated fit to $d=-2$
get very close to the tension of the string by which hadrons can crudely
be fitted, and which were the historically starting point of string theory.

\begin{figure}
  \includegraphics[scale=0.7]{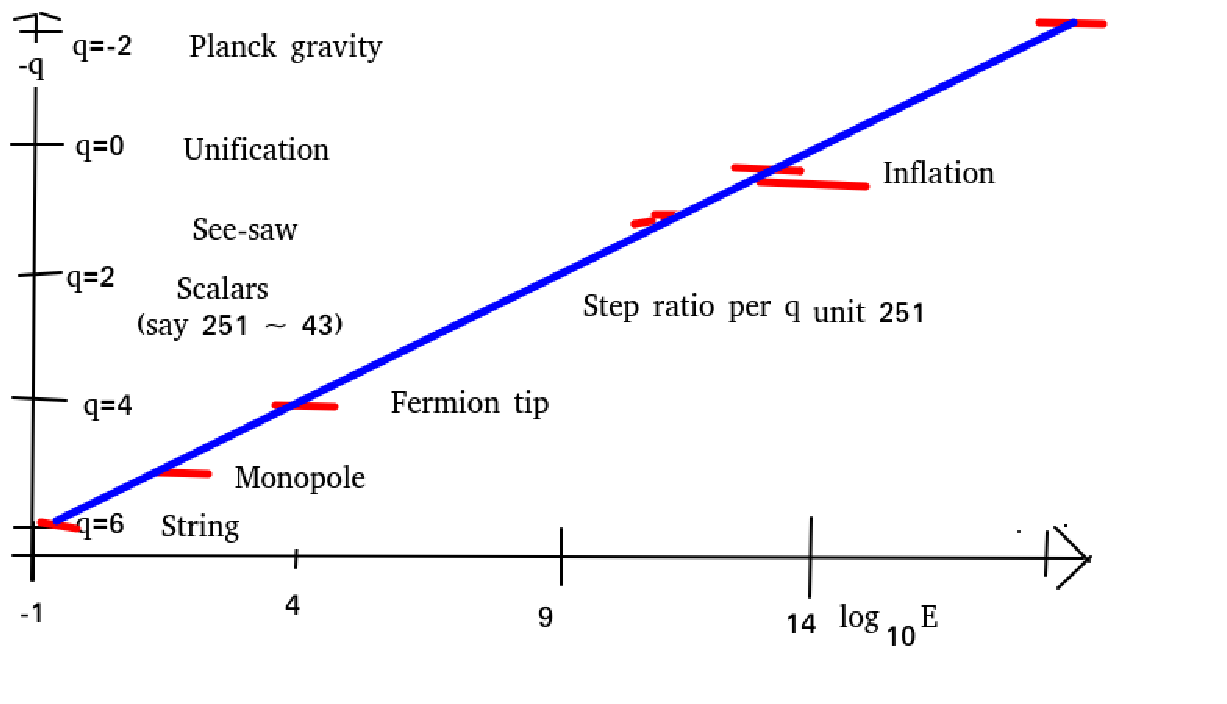}
  \caption{\label{all} This figure taken from \cite{RSR} has the
    logarithm in basis 10 of the energy scales on the abscissa, and the
    intger $q=4-n$, where $n$ is esssentially the power of $1/a$ relevant for
    the scale. Remark that we have $-q$ as the variable pointing upwards.
    The names we have given to the scales are attached.}
    \end{figure}

A plot with the extended system of scales is found in \cite{RSR} and
is reproduced in the figure \ref{all}.

\section{Approximate GUT SU(5)}
\label{sII}
\begin{center}
  {part II: {\huge \bf Approximate SU(5), Fine Structure Constants}} 
  \end{center}
  
{\bf Abstract for Second  part: ``Approximate $SU(5)$, Fine structure constants''}
  \begin{abstract}
    We suggest a model with a  physical lattice for the gauge groups in the
    Standard Model with link variables taking values in the
    according to O´Raifeartaigh Standard Model{\bf group}, $S(U(2)\times U(3))$,
    and it is so similar to $SU(5)$, that in what we can call the classical
    approximation, it gives the same ratios between the three fine structure
    constants. But including quantum fluctuations\cite{tadpole, Niyazi,Vege}
    we get deviation
    from the GUT prediction, becuase there is not true $SU(5)$ and thus
    the true $SU(5)$ quantum fluctuations are lacking, unless they belong
    to the Standard-Model-group\cite{OR}. The remarkable thing is, that
    apart from
      just a factor 3 the deviations caused by these quantum fluctuations
      reproduce within uncertainties in the very accurately measured
      finestructure constants fit the data. The factor 3 we seek to
      explain by postulating that the truly existing lattice
      is really lying in three layers (as if we had three copies of the
      Standard model group, i.e. $SMG\times SMg \times SMG$). 
      \end{abstract}



\subsection{Introduction to approximate SU(5)}
\begin{frame}
  
  {\bf  Fit the three Fine Structure Constants in the Standard Model with
  three Parameters, Derivable in Our Theory.}
  
  Shall fit with\cite{AppSU5}
  \begin{itemize}
  \item $q = \hbox{``number of families''}*\pi/2$.
  \item $\alpha_{5\; uncorrected} = \alpha_{5 \; critical}$ (unifying
    coupling (ours)).
  \item We shall put our replacement for unification energy scale
    $\mu_U$ into aline of four different energy scales, fitting
    a line in the logarithm of the energy scale versus dimension of
    related couplings.
  \end{itemize}
\end{frame}

\begin{frame}

  {\bf Relation to First Part above.: Several ``Fundamental'' Scales, Their
    logarithms Fitted
    on a Line as function of Dimension of the Coefficient in Lagrangian Term
    Related}
  
  Since our replacement for the unification coupling scale is even more
  deviating from the Planck scale than more popular unifications with
  susy\cite{susy},
  we give up that the various `` fundamental scales found, see saw, unification
  (or approximate unification) and Planck scale, should be at the same energy.
  Rather we allow them to vary in a systematic way with the dimensionality
  of the related coefficients in the Lagrangian in the quantum field theory.
\end{frame}
\begin{frame}
  
  
  We interpret this fitting with a model of a truly existing lattice (probably
  irregular) which is fluctuating both in size and local shape, in a way
  corresponding to a fluctuation in the reparametrization gauge of general
  relativity. We though assume that it is somehow cut off so that the
  distribution of the link length say fluctuate on a logarithmic scale
  much like a Gaussian distribution in the logarithm.
\end{frame}
\begin{frame}
  
  When one asks for
  different powers of the link-length for different purposes or different
  types of interactions, one gets the dominant link length to be somewhat
  different. This gives different scales for different purposes or
  lagrange terms: Planck scale, Unification scale, See saw scale,and
  then a scale related to the fermion masses (to be explained).
  
  \end{frame}

  
  {\bf Crossing in one point  of Minimal $SU(5)$ Running (inverse) Fine
    structue constants not
    perfect.}
\begin{figure}\label{fig1}
  \includegraphics[scale=0.8]{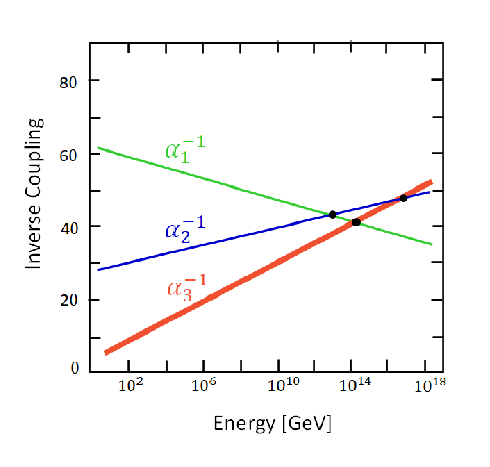}
  \caption{This is the usual graph representing the three Standard Model
    inverse fine structure constants with the $\alpha_1^{-1}$ being in the
    notation suitable for $SU(5)$, meaning it is 3/5 times the natural
    normalization, $\alpha_{1\; SU(5)}^{-1}$ =$ 3/5 *\alpha_{1\;SM}^{-1}$ =$
    3/5*\alpha_{EM}^{-1} cos^2\Theta_W$. The vertical thin line at the energy
    scale
    $\mu_U = 5*10^{13} GeV$ points out ``our unified scale'', which is as
    can be seen not really unifying the couplings, but rather is the scale
    where the ratio of the two independent differences, $\alpha_2^{-1}
    -\alpha_{1\; SU(5)}^{-1}$ and $\alpha_{1\; SU(5)}^{-1} - \alpha_3^{-1}$ have
    just the ratio 2/3 as our model predicts at the `` our unification scale''.
    One may note, that this ``our unified scale'' is actually very close to,
    where the three inverse couplings are nearest to each other, and in that
    sense an ``approximate'' unification scale.}
\end{figure}

    \begin{frame}

      \begin{center}
      {\bf Our prediction of Deviation from $SU(5)$.}
\end{center}

\begin{figure}\label{fig2}
  \includegraphics[scale=0.9]{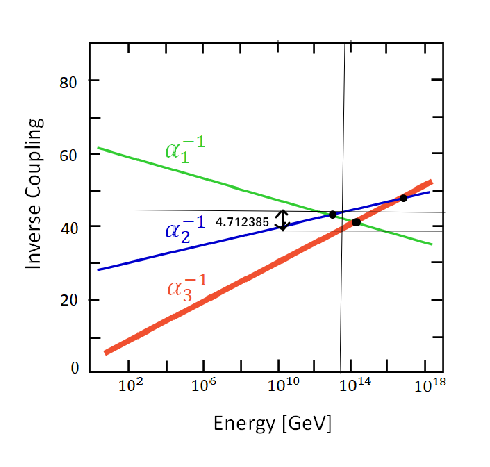}
  \caption{Same as figure 1, but now with our prediction inserted, marked as
    the number $4.712385 = 3*\pi/2$, which is predicted to be at the ``our
    unified scale'' the difference $1/\alpha_2 -1/\alpha_3$. Our prediction
    is, that just at horizontal thin black line, at $5*10^{13} GeV$,
    corresponding
    to the scale $\mu_U$, given by our fitted to the green line crossing point
    dividing the region between the blue and the red in the ratio 2 to 3, we
    shall have
    the difference in ordinate between the red and the blue crossing points
  with the vertical black being $3\pi/2$.}
\end{figure}
    \end{frame}

    \begin{frame}
      
      {\bf Caption for figure 2: Our prediction of deviation form minimal
        $SU(5)$}
      
      \end{frame}

    \begin{frame}
      
      {\bf Our  formulas to be fitted:}
      
      (see \cite{AppSU5} for the details and argument)
      
      The three standard model fine structure constants (inverted):
      
\begin{eqnarray}
  \frac{1}{\alpha_{1 \; SU(5)}(\mu_U)}&=& \frac{1}{\alpha_{5\; uncor.}}- 11/5 *q\\
  \frac{1}{\alpha_{2}}(\mu_U)&=& \frac{1}{\alpha_{5\; uncor.}}- 9/5 *q\\
    \frac{1}{\alpha_{3}(\mu_U)}&=& \frac{1}{\alpha_{5\; uncor.}}- 14/5 *q,
\end{eqnarray}
where the one parameter $\frac{1}{\alpha_{5\; uncor.}}$, is essentially the
unified coupling, although we do not have unification proper.

The deviations come as quantumcorrections on the lattice - the fluctuation of
the link-variables - sometimes refered to as ``tachyon''\cite{tachyon, Niyazi}.
    \end{frame}
    \begin{frame}
      
      {\bf We work with two related `` unified'' couplings,
        $\alpha_{5\; uncor.}$ and $\alpha_{5 \; cor.}$}
        
\begin{eqnarray}
  \frac{1}{\alpha_{5 \; cor.}} &=& \frac{1}{\alpha_{5\; uncor.}}- 24/5*q.
\end{eqnarray}

  The other parameter $q$ we believe to have calculated in our model with
  its 3 families of fermions and in a Wilson lattice in a lowest order
  approximation:
  \begin{eqnarray}
    q&=& ``\# families''*\pi/2 = 3*\pi/2 = 4.712385.
  \end{eqnarray}
    \end{frame}

    \begin{frame}
      
      {\bf Our formulas in ``corrected form'':}
      
  Using this notation we could equally well use the formulation
  \begin{eqnarray}
    \frac{1}{\alpha_{1\; SU(5)}(\mu_U)}&=& \frac{1}{\alpha_{5\; cor.}}+ 13/5 *q\\
       \frac{1}{\alpha_{2}(\mu_U)}&=& \frac{1}{\alpha_{5\; cor.}}+ 3 *q\\
       \frac{1}{\alpha_{3}(\mu_U)}&=& \frac{1}{\alpha_{5\; cor.}}+ 2 *q.
    \end{eqnarray}

    \end{frame}

    \begin{frame}
      
      \begin{center}
      {\bf Table of Fitting the Three parameters}
      \end{center}
      
  \begin{adjustbox}{max width =\textwidth} 
  \begin{tabular}{|c|c|c|c|c|c|}
    \hline
    Parameter& Formula & From $\alpha$'s & Theory & Deviation&Section\\
    \hline
    q &q=$1/\alpha_2(\mu_U) -1/\alpha_3(\mu_U)$ &4.618&  4.712385
    &-0.094$\pm$0.05  &\ref{ps3},
    \ref{dbausu5}\\
    \hline
    $1/\alpha_{5 \; uncor.}(\mu_U)$ &see above& 51.705&45.927&5.778$\pm$ 3.5
    &\ref{ps10}\\
    \hline
    $\ln(\frac{\mu_U}{M_Z})$ &$\ln(\frac{\mu_U}{m_t}) =\frac{2}{3}*
    \ln(\frac{E_{Pl\; red}}{M_t})$  & 
    27.04 & 24.76 &
    2.28$\pm$ 1 
    &\ref{scale}\\
    &&&&or 0.02&\\
    \hline
\end{tabular}
  \end{adjustbox}

  \vspace{3mm}
  
  In the third line we now replace the top mass $m_t$ with a mass
  value gotten by extrapolating from the whole spectrum of quarks and leptons,
  which is about $10 TeV$ and the agreement got very good indeed.
    \end{frame}
    
    \subsection{Model}
    
    \begin{frame}
      
      \begin{center}
        {\bf We assume ANTI-GUT: Diagonal subgroup breaking}\\
        \cite{confusionetal,Bled8,Picek,RDrel1,RDrel2,RDrel3,RDrel4,Don137} 
      \end{center}
      
  \begin{eqnarray}
      G_{full}&=& SMG\times SMG \times SMG\\
      \hbox{where } SMG &=& S(U(1)\times U(3))\\
      &=&({\bf R}\times SU(2) \times SU(3) )/{\bf Z}_{app}\\
      \hbox{where }&&\nonumber\\
           {\bf Z}_{app}&=& \left \{ (r, U_2, U_3) |\right.\nonumber\\
          && \left. \exists
      n \in {\bf Z}
      [(r,U_2,U_3) =\left ( 2\pi,-{\bf 1}, \exp(i2\pi/3) {\bf 1}\right )^n]
      \right \}\nonumber
      \end{eqnarray}
  \begin{eqnarray}
    SMG_{as \; observed} &=& \{ (g,g,g)|g\in SMG\}\subset G_{full}
    \end{eqnarray}
      \end{frame}
    \begin{frame}
      
      {\bf Action: Trace of in ${\bf 5}\times {\bf 5}$ imbedding of SMG}

      It is our crucial assumption that we have a lattice theory with
      plaquette-action given proportional to  the trace of the representative
      of the plaquette group element $U_{Pl}(\Box)$ in the/a ``smallest''
      representation\cite{sr,srcim4,srKorfu14} - taken here as the representation in the five-plet
      $SU(5)$ representation ${\bf 5}$:
      
      \begin{eqnarray}
  \rho(U_{pl}(\Box) &:& {\bf 5}\rightarrow {\bf 5}\\
  \hbox{or }\rho(U_{pl}(\Box))& \in & UnitaryMarix(5 \times 5)
\end{eqnarray}

We have once pointed out that the very standard model group $SMG$ is
selected\cite{sr, srKorfu14, srcim4} as the one having with appropriate definition the smallest
relative to the group faithful representation.
      \end{frame}
    
    To get a predition for the unified coupling at the approximate unification
    scale we make the assumption that somehow the coupling constants at
    this scale becomes just couplings which are``critical''in the sense of
    seperating
    typically several phases of the lattice theory. We call this assumption
    ``multiple point criticallity principle''(MPP)\cite{RDrel1,RDrel2,RDrel3,
    RDrel4, Don93,Don137}.
    
    \subsection{Fitting}
    
    \begin{frame}
      
      {\bf We predict two differences between $1/\alpha$s in Absolute number}

      Taking our $q=3\pi/2$ as just given in our model, and we predict the
      differences from a to be fitted ``unified'' inverted coupling
      $1/\alpha_{5\; uncor.}$ at a to be fitted
      ``unification scale'' $\mu_U$, we really only provide {\bf one
        predicted parameter at first}. Really we predict the two independent
      differences, say $1/\alpha_2 - 1/\alpha_3 = q$ and
      $1/\alpha_2 - 1/\alpha_{1\; SU(5)} = 2/5*q$ at the ``unification scale''.

      E.g. select the scale by having the ratio of the two differences the
      predicted one; then the absolute size is a true prediction.

      We got $q=4.6 $ by the fine structure constant data and the
      $3\pi/2 = 4.712$.
      
      \end{frame}
    \begin{frame}

\vspace{8mm}
      
      {\bf Inverse Fine structure constants at the $\mu_U$-scale}\\

      
      \includegraphics[scale=0.7]{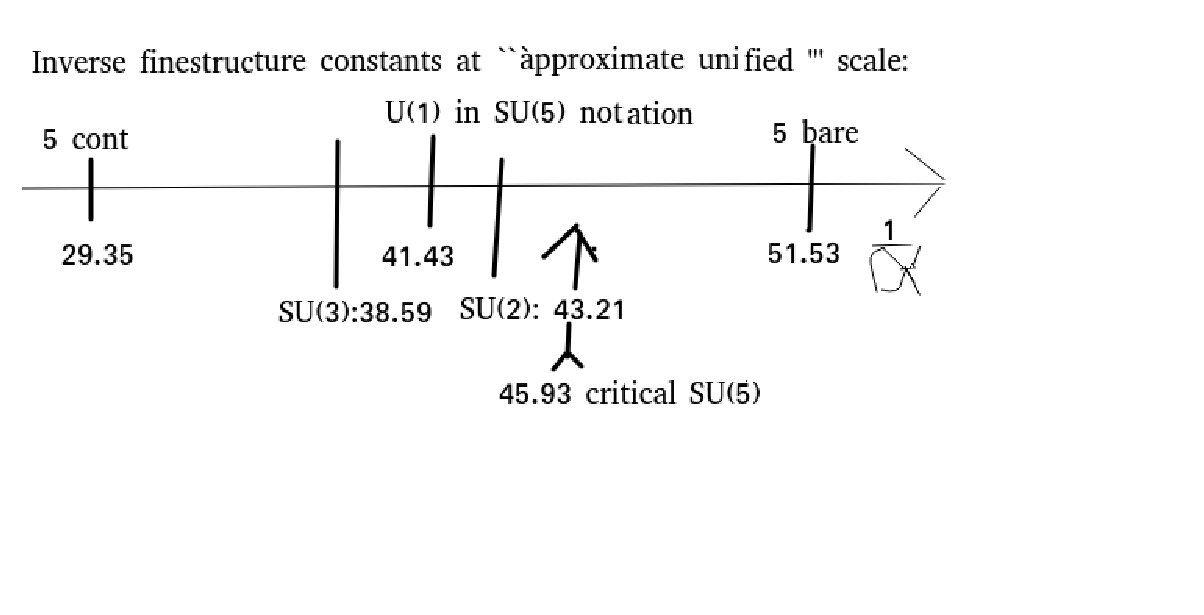}
    \end{frame}
    \begin{frame}
      
      \begin{center}
        {\bf Figure Caption about Critical Coupling}\\
        \cite{crit,LRN,LRNwf,RDrel2,RDrel1,Picek}
      \end{center}
      
Explaing figure:The axis is the axis of inverse fine structure constants;
      The group names U(1), SU(2) and Su(3) are the by renormalization group
      to the replacement of unification scale $\mu_U$ extrapolated expeimental
      inverse fine structure constants for these groups respectively. The two
      $SU(5)$ inverse fine structure constants are repectively with and
      without the quantum fluctuation contribution.
      
      \end{frame}
    \begin{frame}
      
      {\bf Helping Approximations to justify Critical Coupling}

      To justify that the above figure implies that the unified coupling
      represented by the inverse fine structure constant has indeed
      the critical value (for a phase transition, presumably
      between confinement or not) we make use of the following three
      approximations/assumptions:(see next slide)
    \end{frame}
    \begin{frame}
      
      {\bf The approximations or assumtions}
      
      \begin{itemize}
      \item The critical couplings for a true $SU(5)$ lattice theory
        and for the Standard Modelgroup\cite{OR, sr,srKorfu14, srcim4 } deviate
        only little, because the
        standard model group can be considered an attenuation of the
        SU(5) one.
      \item  We can trust a rather simple formula \cite{LR, LRSU5} for the
        critical couplings for the SU(N) groups,
        \begin{eqnarray}
          \frac{1}{\alpha_N} &=& \frac{N}{2}\sqrt{\frac{N+1}{N-1}}
          \alpha_{U(1) crit}^{-1}\\
            \hbox{where } \alpha_{U(1)crit}^{-1}&=& 0.2 \pm 0.015
        \end{eqnarray}
        found in an article with Laperashvili and others\cite{LRSU5}.
      \item The critical coupling for the Standard Model group
        $S(U(3)\times U(2))$ should be compared to couplings with equally
        many quantum fluctuation contributions as it has itself.
      \end{itemize}

      And then the assumption of the model which is not only an approximation:

      We compare the `` unified couplings'' not to simply the critical one
      but the by a factor 3 weakened one, so that we have multiplied
      the critical inverse finestructure constant for the Standard Model
      group by 3, to compareit with  the unified couplings.(the 3 is again
      the number of families)
    \end{frame}

    \begin{frame}
      
      {\bf A reference to Larisa et al.\cite{LRSU5}}
      
    \end{frame}
    \section{Gravity Problem; Return to First Part again}
    \label{sp}
    \begin{frame}
      
  
      {\bf \huge Return to Part I: on the scales in fluctuating lattice model}

      Really I believe that gauge symmetries could be due to very huge
      fluctuations in thosedegreesof freedom which arethe guage-monters.

      If a lattice were connected to a coordinate system in general relativity,
      but the guage not fixed but allowed to fluctuate\cite{NNF, }, we should get
      a lattice fluctuating relative to what we would consider the fixed
      geometry.
      
    \end{frame}
\begin{frame}
  
    {\bf Fluctuating Lattice Imposed by General relativity}

    \includegraphics[scale=0.6]{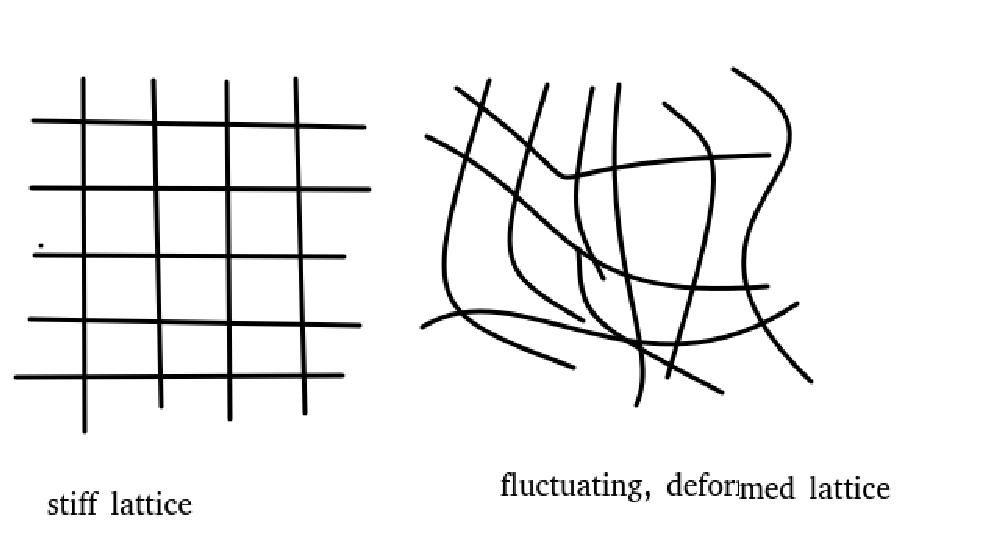}
    
\end{frame}    
\begin{frame}
  
  \begin{center}
    {\bf Contributions as function of ln scale in fluctuating lattice:}
    \end{center}
  \begin{figure}
  \includegraphics[scale=0.7]{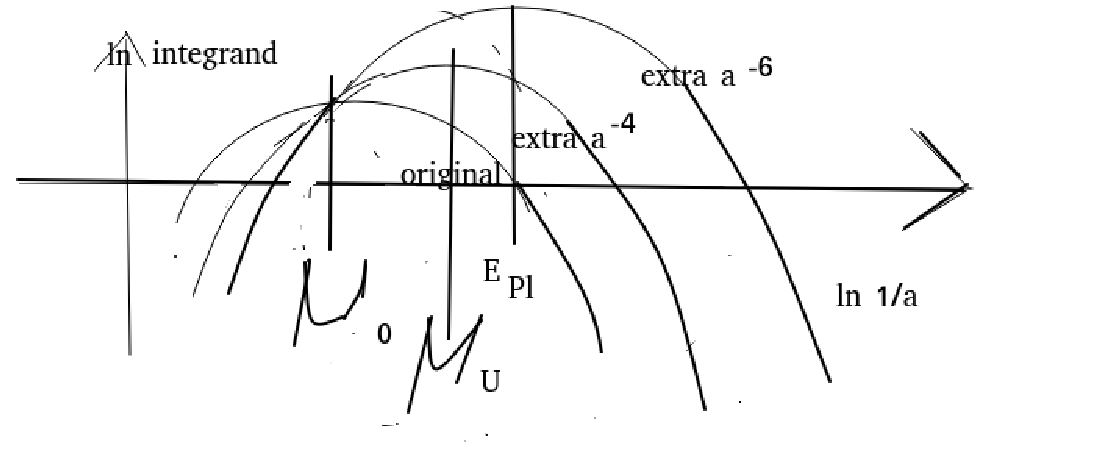}
   \caption{\label{shifted} {\bf Shifted Scales Depending on Weighting
        with $(1/a)^n$ weight factor}:
  On the abscissa we have the logarithm $\ln(1/a)$ of energy $1/a$, and
  for several cases of weighting with powers of the $1/a$ we have on the
  ordinate the logarithm of  the contribtuion density to the average of
  these powers of $1/a$. In the Gauss distribution assumption, which we use
  the logarithm of the distribution density as function of the
  $\ln(1/a)$ is a parbola (pointing dowward) and for the various powers
  of $1/a$ shown the weighted distribution becomes again a now
  displaced parabola. It is the displacement component along the
  abscissa here, which represents the change in effective energy scale,
  which is the central effect dicussed in the first part of this paper.}
  \end{figure}
  
    \end{frame}

  
  \begin{center}
  \end{center}
  
\begin{figure}
  \includegraphics[scale=0.6]{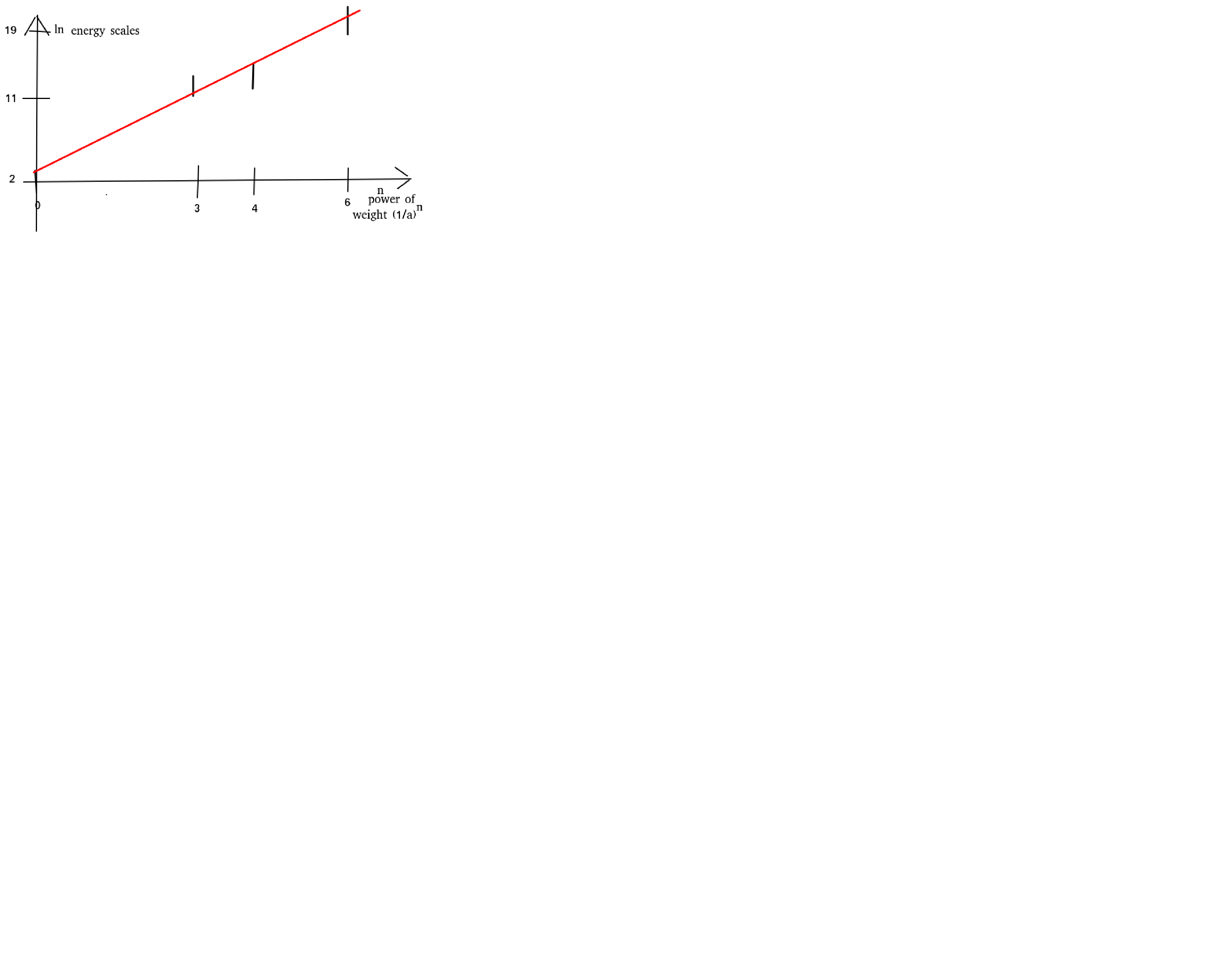}
  
  \caption{\label{wtop} Here we attempted to fit our first four energy
    scales with the ``fermion tip'' scale approximated by just the
    top mass $m_t$, because of course after all the top quark is the heaviest,
    so it is approximately the tip scale. On the abscissa we have the
    energy scale characteristic integer $n$ essentially denoting which power
    $(1/a)^n$ is relevant for the energy scale in question. The ordinate
    is the logarithm with basis 10 of the energy scale. We use base 10, because
    it makes it easier to relate to our ten-power notation for the energy
    scales.}
\end{figure}
\begin{frame}
  
  \begin{figure}
    \includegraphics[scale=0.6]{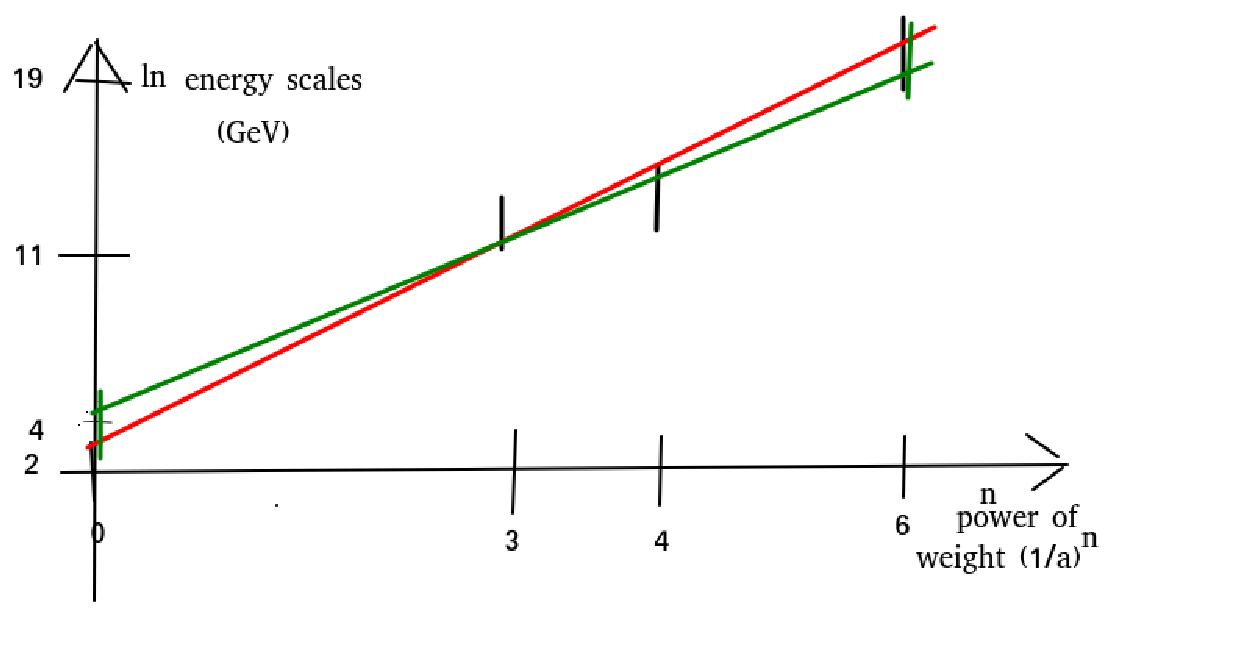}
    \caption{\label{wmmnl} However, taking for the ``fermion tip''
      scale rather than just top mass $m_t$ the value gotten by the
      extrapolation $m_{mnl}$ as seen in figure \ref{extrapolation} and in the
      tables for respectively quarkmasses and leptonmasses not long below.
      The fitting is a bit better than for using the top-mass simply.}
   \end{figure}
  
\end{frame}
\begin{frame}
  
  \begin{center}
      {\bf The plot of scales versus weighting power $n$}
  \end{center}
  
      We present 4 energy scales of physical interest together with
      lattice link size $a$ dependent factor comming into the expression
      in the action or Lagrangian relevant for the scale in question. It is
      essentially the dimension of the termin the Lagrangian density without
      counting the coefficient (so it is rather trivially related to this
      coefficient).  We took:
      
      \begin{itemize}
      \item {{\bf 0} $(1/a)^0$ } This scale is the scale of maximal number
        of ``active''/effectively massless families. (Needs more explanation
        below.). Below extrapolation $\sim 10 TeV$.
        
      \item{{\bf 3} $(1/a)^3$ } The see-saw neutrino mass scale

      \item{{\bf 4} $(1/a)^4$} The our ``unification scale'',at which
        the Yang Mills theories are supposed to be given by the truly existing
        fluctuating lattice.
      \item{{\bf 6} $(1/a)^6$ } The Planck scale, related to the
        Einstein-Hilbert-action. 
        \end{itemize}
    \end{frame}

      \begin{figure}
        \includegraphics[scale=0.6]{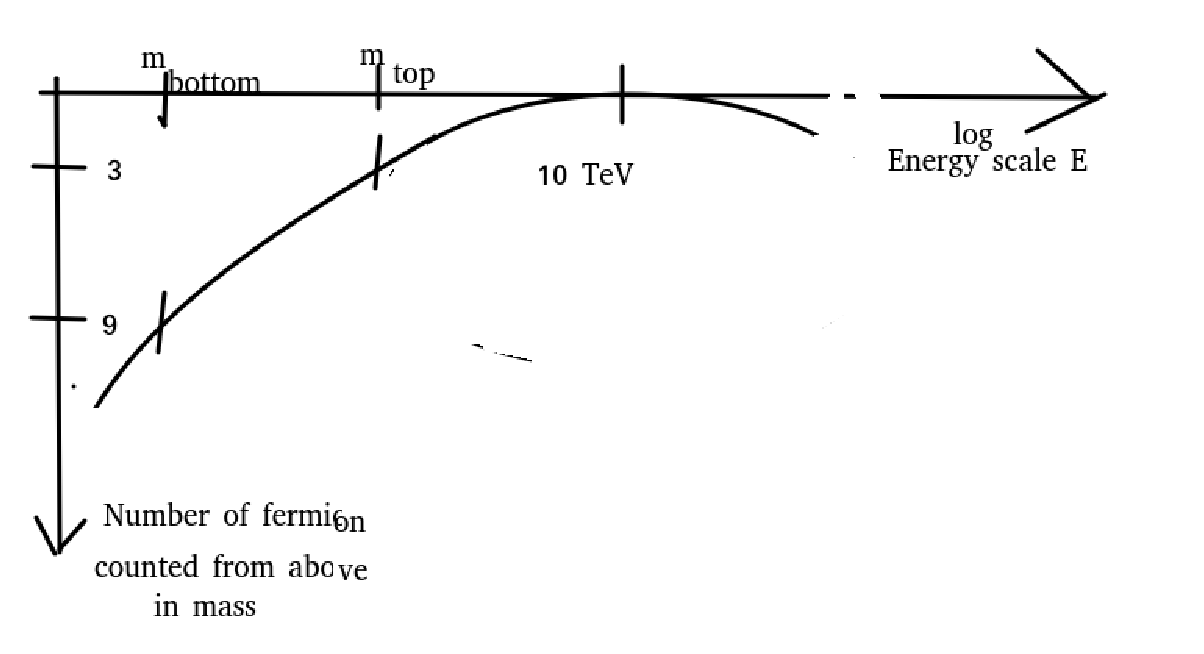}
        \caption{\label{extrapolation}   {\bf Fitting Weyl Fermion
            masses by
          number after mass by Philosophy of Fluctuating Lattice:}
         In order to make an as good as possible
      extrapolation to where the series of masses of the fermions in the
      Standard Model we should assume how this series of masses is related to
      some ansatz formula for their density on say the logarithm of mass axis.
      Inspired by our ideas of a fluctuating lattice with a Gaussian conbined
      with a connection of the number of fermions with relative to that scale
      negligible mass being proportional to the layer density at the scale,
      we suggested: The number of a fermion counted from the heaviest, the next
      heaviest, etc. should be approximately proportional to the square
      of the distance in logarithm from the ``extrapolated tip'' of the
      mass region, which we call $m_{mnl}$ to the massoof the fermion. On this
      plot we see as function of the logarithm of masses the number in the
      series counted from the heaviest down according to mass. We have let
      the axis denoting
      the number in the series point down. So the curve is to be fitted by
      a (half) parabla.}
      \end{figure}

    \begin{frame}

      \begin{center}
      {\bf Quark for $m_{mnl}= 10^4GeV$ }
      \end{center}
      
 $m_{mnl}$ = ``maximum number of layers'' is the energy scale at which
      density
      of the distributuion of inverse lattice sizes $1/a$ is the biggest.

      We fit to this density being proportional to the number of Weyl fermions
      relative to the scale being light/massless. Since the last column
      $diff^2/n$ fits a constant 1.12 to about 0.1 we have good fit for the
      10 TeV.
      
\begin{adjustbox}{width=\textwidth, center}
\begin{tabular}{|c|c|c|c|c|c|c|}
  \hline
  Name& number $n$&Mass $m$&$log_{10\; GeV}m$&$diff$=$4-log m$&
  $diff^2$&$diff^2/n$\\
  \hline
  top&3$\pm$ 1&172.76  $\pm$ 0.3 GeV&2.2374$\pm$ 0.0008&1.7626&3.1066$\pm$
  0.003&1.0355
  $\pm$ 0.001 $\pm $0.4\\
  \hline
  bottom&9$\pm$ 0.3& 4.18$\pm$ 0.0079GeV& 0.6212$\pm$0.001&
  3.3788&11.416$\pm$ 0.01&1.268$\pm$ 0.001$\pm$0.03 \\
  \hline
  charm&17 or 15&1.27$\pm$ 0.02&0.10382$\pm$ 0.009&3.8962&15.180$\pm$ 0.07&
  0.893 $\pm$ 0.004 $\pm$ 0.06\\
  \hline
  strange&25 or 23&0.095$\pm$0.006 GeV& -1.0223$\pm$0.003 &5.0223&25.223
  $\pm$0.03&1.009$\pm$0.001$\pm$0.1\\
  \hline
   down& 31&4.79$\pm$ 0.16 MeV&-2.3197$\pm$0.01 &6.3197&39.939$\pm$ 0.06&1.288
   $\pm$ 0.002 \\
   \hline
  up&37&2.01$\pm$ 0.14 MeV&-2.6968$\pm$0.03& 6.6968&44.847$\pm$ 0.4&1.212$\pm$
  0.01\\
  \hline
\end{tabular}
\end{adjustbox}

\end{frame}
    \begin{frame}

      \begin{center}
      {\bf Leptons for $m_{mnl}= 10^4GeV$ }
      \end{center}
      
      $m_{mnl}$ = ``maximum number of layers'' is the energy scale at which
      density
      of the distributuion of inverse lattice sizes $1/a$ is the biggest.

      We fit to this density being proportional to the number of Weyl fermions
      relative to the scale being light/massless. Since the last column
      $diff^2/n$ fits a constant 1.19 to about 0.1 we have good fit for the
      10 TeV.
      
\begin{adjustbox}{width=\textwidth, center}
\begin{tabular}{|c|c|c|c|c|c|c|}
  \hline
  Name& number $n$&Mass $m$&$log_{10\; GeV}m$&$diff$=$4-log m$&
  $diff^2$&$diff^2/n$\\
  \hline
  $\tau$&13 or 19&1.77686$\pm$0.00012&0.2496$\pm$ 0.00003&3.7503&14.065$\pm$
  0.0003&1.082
  $\pm$ 0.00002 $\pm $0.4\\
  \hline
  mu&21 or 27&105.6583745$\pm$ $2.4*10^{-6}$MeV&$-0.9761...\pm$ $10^{-8}$&4.9761&
  24.761$\pm$ $10^{-7}$&1.179$\pm$ $4*10^{-9}$ $\pm$ 0.3\\
  \hline
  electron&41&0.51099895069$\pm$$1.6*10^{-10}$& -3.2915$\pm$ $4*10^{-10}$&
  7.2916&53.167$\pm$ $10^{-8}$&1.297$\pm$ $10^{-11}$\\
  \hline
\end{tabular}
\end{adjustbox}

\end{frame}

    \begin{frame}
      
      {\bf Explaining the tables fitting Fermion Masses to Fluctuating Lattice}

      In the two foregoing tables - one for quarks, the second one for
      the charged leptons - youhave in first column the name of the fermion,
      then its number in the series of fermions counted as Weyl fermions
      and after mass, the heaviest first then the lighter and lighter ones.
      A quark flavour corresponds to two Weyl per particle and it has three
      colors,so there is under each flavour 6 Weyl and werepresent a flavour
      by the midle one of these 6. So the top quark gets the representative
      number $n=3$ ( the midlebetween 0 and 6. We use logarithmic scale and
      care
      for the logarithm - we use log of basis 10 for slightly easier
      calculation - of the ratio of the fermion mass to the scale we test with
      as $m_{mld}$ = `` maximum layerdensity point on the energy scale''.
    \end{frame}
    \begin{frame}
      
      
      Because our best fit  $m_{mld}= 10^4 GeV = 10 TeV$ the log of it
      is just the 4 in the column $4-log_{10}m$ (= diff).Since we want to
      fit the
      number of layers as a square function  of the logarithm of the masses,
      we shall
      square what in the table is called $diff$ and which is just the log of
      theratio mentioned.

      If the Fermion masses were indeed arranged so as to make the number
      of (Weyl)fermions with mass under a given scale be proportional to the
      a quadratic function in the log dropping down from a maximum as we go
      more and more below
      $m_{mnl}$ point, then the ratio in the last column $diff^2/n$ should
      be constant.

      If we would have liked to fit with a Gaussian of the logarithm
      of the masses, weshould intead of the number $n$ have used
      $\log \frac{45 -n}{45}$, which for the first small $n$ is approximately
      proportional to $n$ itself.(45 isthe number of Weyl particles in SM).

    \end{frame}
    
    \subsection{Seesaw}
    
    \begin{frame}
      
      \begin{center}
        {\bf What to take for the Seesaw neutrino scale?}\\
        \cite{Takanishi, Takanishi1, qarketcmasses, King, Mohapatra,Grimus, Davidson}
\end{center}
      
\begin{adjustbox}{width = \textwidth}
      \begin{tabular}{|l|c|c|}
        \hline
        Name&Seesaw-scale&Comments\\
        \hline
        Steven King&$3.9*10^{10}GeV$&lowest mass;susy\\
        \hline
        Grimus and Lavoura& $10^{11} GeV$ &\\
        \hline
        Davidson and Ibarra& $\ge 10^9 GeV$&\\
        \hline
        ``statisic'' (my own)& $1.4*10^9 GeV$&\\
        \hline
        Mohapatra& $10^{14}$ to $10^{15}$ GeV& very crude guess\\
        \hline
        Modernized Takanishi and me(own)& $1.2*10^{15}GeV$&\\
        \hline
        \hline
        Average of most trustable& $10^{11}GeV$&\\
        \hline
      \end{tabular}
      \end{adjustbox}
      \end{frame}

\section{Conclusion}

\subsection{Conclusion for part I: Several scales fit on a line}
      \begin{itemize}
      \item We related four {\bf different} physical/``fundamental''
        scales by a line relating the energy scale to their dimention
        of the related Lagrange density term.

        {\bf 2.5 orders of magnitude per dimension of the Lagrange term
          coefficient}.

        The four scales
        are:
        \begin{itemize}
        \item A scale relatedto the fermion mass distribution of
          formal dimension of coupling $[GeV^4]$.
        \item The See saw scale, coupling dimension $[GeV]$,
        \item Approximate  Grand unification scale, coupling dimension
          $[1]$,
          \item Planck i.e. gravity scale, coupling dimension $[GeV^{-2}]$ 
          \end{itemize}
        \end{itemize}

 \subsection{Conclusion for Part II on Approximate SU(5) GUT}
    

      We had a successful agreement with the values of the fine structure
      constants in a minimal (i.e. no susy!) approximate GUT SU(5)
      ``unification'' at the scale $\mu_U = 5.13*10^{13} GeV$ (compared
      susy-models a very lowenergy scale, but not far at all from the
      scale needed for see-saw neutrinoes to fit the neutrino oscillations).
      
      

      We have three parameters {\bf predicted by our theory}:
      \begin{itemize}
      \item $q=3*\pi/2$ is a parameter going into the {\bf deviation} from
        full GUT-SU(5).
      \item The replacement for the unification coupling
        $\alpha_{5\; uncor.} $ or $\alpha_{5\; cor.}$,whichever one of them we
        want to
        think of, or rather the thirds of one of them, should correpond to
        the critical value,in the sense that
        it should be at borderline of two phases of the vacuum.
      \item The replacement for the unification scale $\mu_U$ goes into
        a series of ``fundamentalscales'' fitted on a line.
        
          \end{itemize}

    \subsection{Later developments}
    But since the conference virtually in Bledwe added:
    \begin{center}
      {\bf Added later Scales:}
    \end{center}

    \begin{itemize}
    \item ``scalars'' is a scaleonly in my phantasy at which there
      should be a lot of scalar boson masses. associated with that
      presumably also some non-zero expectation values breaking
      dreamt about symmetries yet to be discovered spontaneously. Such
      breakings of symmetries by the ratio of this ``scalars''scale to the
      ``see saw'' scale which is the 1/250 could be the weak breaking
      reponsible for the small hierarchy problem, that the ratio between
      the fermion masses in the Standard Model typically are large by
      by factors not so different from 250.
    \item A ``monopole'' scale alsodreamt up of masses of
      presumably bound states of some monopoles for the standard model
      groupbeing confined by their SU(3) part of the monopolic charge.
      Actually a candidate for having found them is a dimuon-resonance,
      which is about the only new physics surviving from the LHC\cite{dimuon},
      also
      by reanalysis seen in LEP\cite{Heister}.
    \item A string theory with the energy scale given by our fluctuating
      lattice turns out to agree surprizingly well with the  string theory for
      hadrons, that were historically the first string theory application.
    \item The energy scale for ``2-branes'' with Goto (Nambu) action
      would get from our fluctuating lattice a scale of tension not
      violently different from what Coin Froggatt and I get from
      phenomenological fitting of dark matter as pearls of new vacuum
      encapsulated by the ``2-branes''. 
    \end{itemize}

    It may be needed to admit, that although we now added a scale
    ``scalars'' at which there should be lots of scalar particle
    masses, the only fundamental scalar physicists found so far, the Higgs,
    does not at all have the mass of these phantasy scalars. We would have to
    excuse it by supposing it is fine tuned some way that dominate over
    the prediction of our scheme. Similarly the cosmological constant could
    get a predicted order of magnitude in our scheme, but also that
    is fine tuned in a way making it not at all agree with our scheme.
    My own guess is that these fine tuned quantities would rather find their
    solution in schemes similar to our complex action theory \cite{CAT2, CAT,
      Frampton}, but at least they escape our prent article system.
    In \cite{Frampton} I really do argue for a very small cosmolgical constant
    comming out of this type of complex action theories.
    
    \section*{Acknowledgement}
    The author thanks the Niels Bohr Institute for status as
    emeritus including a working room and participants at the virtual Bled
    Conference
    for discussions. Konstantinos Anagnostopoulos is thanked for the reference
    to Senjanovic \cite{Senjanovic}. Astri Kleppe is thanked for having removed
    printing mistakes in large amounts.


     \begin{frame}

       \end{frame}



\end{document}